\documentclass[conference]{IEEEtran}
\IEEEoverridecommandlockouts
\usepackage{cite}
\usepackage{amsmath,amssymb,amsfonts}
\usepackage{algorithm}
\usepackage{algorithmic}
\usepackage{graphicx}
\usepackage{textcomp}
\usepackage{xcolor}
\usepackage[a4paper, total={184mm,239mm}]{geometry}

\usepackage{color}

\def\BibTeX{{\rm B\kern-.05em{\sc i\kern-.025em b}\kern-.08em
    T\kern-.1667em\lower.7ex\hbox{E}\kern-.125emX}}
\begin{document}

\title{NTTSuite: Number Theoretic Transform Benchmarks for Accelerating Encrypted Computation}

\makeatletter
\newcommand{\linebreakand}{%
  \end{@IEEEauthorhalign}
  \hfill\mbox{}\par
  \mbox{}\hfill\begin{@IEEEauthorhalign}
}
\makeatother

\author{
\IEEEauthorblockN{Juran Ding}
\IEEEauthorblockA{
\textit{New York University}\\
jd4691@nyu.edu}
\and
\IEEEauthorblockN{Yuanzhe Liu}
\IEEEauthorblockA{
\textit{New York University}\\
yl7897@nyu.edu}
\and
\IEEEauthorblockN{Lingbin Sun}
\IEEEauthorblockA{
\textit{New York University}\\
ls5382@nyu.edu}
\linebreakand  %
\IEEEauthorblockN{Brandon Reagen}
\IEEEauthorblockA{
\textit{New York University}\\
bjr5@nyu.edu}
}

\maketitle

\begin{abstract}
Privacy concerns have thrust privacy-preserving computation into the spotlight.
Homomorphic encryption (HE) is a cryptographic system that enables computation to occur directly on encrypted data, providing users with strong privacy (and security) guarantees while using the same services they enjoy today unprotected.
While promising, HE has seen little adoption due to extremely high computational overheads, rendering it impractical.
Homomorphic encryption (HE) is a cryptographic system that enables computation to occur directly on encrypted data. In this paper we develop a benchmark suite, named NTTSuite, to enable researchers to better address these overheads by studying the primary source of HE's slowdown: the number theoretic transform (NTT).
NTTSuite constitutes seven unique NTT algorithms with support for CPUs (C++), GPUs (CUDA), and custom hardware (Catapult HLS).
In addition, we propose optimizations to improve the performance of NTT running on FPGAs.
We find our implementation outperforms the state-of-the-art by 30\%.
\end{abstract}

\section{Introduction}

The technology industry has recently begun to re-think user privacy while also dealing with ever-raising security threats.
Legislature, e.g., GDPR~\cite{GDPR}, and large fines, e.g., Facebook's five billion dollar fine~\cite{fb5b}, are reshaping how user data is collected, used, and stored to substantially increase privacy rights.
Attacks, stemming from hardware (e.g., Spectre~\cite{Spectre_Kocher2019} and Meltdown~\cite{meltdownattack}) and software further demand increased security.
Fortunately, there is an emerging computational paradigm, which we refer to as privacy-enhanced computing (PEC), that enables computation to occur directly on encrypted data.
With PECs, users' gain significant benefits in both privacy and security, as all data leaving a local device is always encrypted, while still enjoying the utility of services now fundamental to our daily lives.
One promising PEC technology is homomorphic encryption (HE)~\cite{PEC}.
While HE has the potential to address many privacy and security issues, it incurs extreme performance overheads that severely restrict its practicality.

Homomorphic encryption secures data via lattice-based cryptography.
In modern HE schemes, data are encoded into polynomials with noise.
When performing computations in HE, these polynomials must frequently change representation to improve performance.
This is done via the number theoretic transform (NTT).
The NTT is a variation of the more familiar FFT, and it can be used to reduce polynomial multiplication runtime from $O(n^2)$ to $O(nlogn)$.
Where multiplication can be significantly sped up in the NTT (or evaluation) domain, some functions can only be processed in the native, or coefficient, representation.
In HE, NTT is among the most expensive function, due to the frequency of transformation of polynomial cipher text and the complexity of NTT.
For example, a recent paper profiled a machine learning algorithm running in HE and found that 55\% of the total run time could be attributed to NTT~\cite{cheetah}.

We propose and develop a benchmark suite for studying and optimizing the NTT on accelerated platforms, including GPUs and custom hardware.
We believe a benchmark is necessary for two reasons.
First, the NTT is a complex workload with many different implementations and optimizations for locality and parallelism.
With a standard set of implementations, which serve as references, researchers can focus on the intellectual challenges of optimizing NTT without having to rebuild test and research infrastructure.
Moreover, by implementing many popular NTT varieties we can enable users to compare and contrast which NTTs are most amenable to different hardware platforms and hardware optimizations.
Second, a series of papers on NTT (and HE) accelerators now already exists, this is a trend we expect to continue and grow over the coming years.
Thus having a common set of implementations improves commensurability across research from different groups and facilitates reproducible results.
The strides made in accelerating HE and NTT have already been substantial;
With a shared NTT benchmark, more researchers can work in the area to help achieve the full potential of HE.

In this paper we present and develop NTTSuite\footnote{This work was done while Juran, Yuanzhe, and Lingbin were Master degree students at NYU. This report documents experiences, experiments, and reference code for those interested.}: a collection of reference implementations for standard NTT algorithms to enable performance and efficiency optimizations on accelerated platforms, including GPUs and custom hardware.
NTTSuite constitutes seven core NTT algorithms that highlight the differences in how NTTs are typically implemented, including DIT, DIF, Flat-NTT, Pease, Pease\_nc, Six-step, and Stockham.
While building the benchmarks, we realized an opportunity for a more efficient implementation of the Pease algorithm that elides data copies between stages.
We call this implementation Pease No Copy (Pease\_nc) and include it in NTTSuite for a total of seven benchmarks.
In addition to implementations, we have a testing environment derived from the CPU implementation.
This way, any user can run the CPU (C++) version to get parameter settings and input/outputs to validate accelerator implementations, both for GPU and FPGA.
The core of the benchmark is the accelerator implementations, which we believe are most important, as overcoming the large slowdowns of HE requires custom hardware.
NTTSuite provides verified implementations of all seven benchmarks using Catapult HLS.
With NTTSuite, users can download the code and immediately begin optimizing NTT accelerators with HLS pragmas and code rewriting while comparing hardware results to published results and verifying designs against our test harness.

To demonstrate the utility of NTTSuite, we profile the benchmark on all three back-ends using a range of problem sizes, from 1024 to 16384 points.
The experiments highlight the versatility of the benchmark and provide a set of baseline numbers that researchers can use to improve across different devices and scenarios.
We show that HLS optimizations can be made to perform well using a combination of unrolling, partitioning, and pipelining pragmas matched with the careful selection of SRAM type.
Through experimenting with designs we found that the modular arithmetic was particularly inefficient.
To resolve this, we implement and include in NTTSuite a prior optimization~\cite{heax} for efficient modular reduction.
We find that the optimized functional units substantially outperform the native HLS reduction logic by up to 12$\times$.
Finally, we compare our novel NTT algorithm with HLS optimizations against a recent competitive design, HEAX~\cite{heax}, and demonstrate a 30\% performance improvement while using fewer resources.

This paper makes the following contributions:
\begin{enumerate}
    \item We develop (and release) NTTSuite: a collection of seven NTT algorithms and test harness for CPU, GPU, and custom hardware support.
    \item We develop a novel NTT algorithm designed to perform well in custom hardware, named Pease\_nc.
    \item We optimize, profile, and implement the NTTSuite benchmarks and find our novel NTT algorithm with pragmas and modular reduction optimizations outperforms the current state-of-the-art.
\end{enumerate}

\section{The NTTSuite Benchmarks}
In addition to the textbook DIT and DIF algorithm, there exist other variations of the NTT algorithm tailored to different computing platforms and microarchitectures. NTTSuite supports seven unique NTT algorithms to enable researchers to compare both the algorithmic tradeoffs on different hardwares and select the best fit for their platform.
The seven were chosen to span the tradeoffs in the NTT algorithm design space.
Some algorithms (e.g., Pease) have very regular computational patterns at the expense of data shuffling whereas others are able to do more computing before shuffling data, but the compute patterns are more complex.
In our our evaluation of NTTSuite, we found that an optimized Pease algorithm is suitable to fully utilize unrolling and pipelining in FPGA design due to its unique memory access pattern.

\textbf{NTTSuite}: 
We implement each NTT algorithm in NTTSuite on different computation platforms including C++ for CPUs, CUDA for GPUs, and Catapult HLS for FPGAs.
Details on each NTT algorithms are provided below. 
Our new algorithm, Pease\_nc aims to remove memory copy in the Pease algorithm while allowing us to fully pipeline and make the computation parallel using the same degree of memory partition.
In the evaluation section, we show that based on the time and complexity savings from the reduced data movement, our final algorithm peace\_nocopy results in the best-performing hardware design.
Below we describe NTT and algorithms.

\textbf{Twiddle Factors}:
Considering the Discrete Fourier Transform and Number-Theoretic Transform:
\begin{equation}
\begin{split}
    X_f[k]&=\Sigma^N_{n=0}x_f(n)e^\frac{-j2\pi nk}{N}\\
    X_{ntt}[k]&=\Sigma^N_{n=0}x_{ntt}(n)g^{\frac{P-1}{N}nk}
\end{split}
\end{equation}
in which $k$ is the index in the new representation, $n$ is the index of the original representation, $N$ is the number of data points, and $P$ is a prime. The only difference in the equations is term $e^\frac{-j2\pi nk}{N}$ and $g^{\frac{P-1}{N}nk}$. And from the periodicity of $e^\frac{-j2\pi n}{N}$ and Fermat's little theorem:
\begin{equation}
\begin{split}
        e^\frac{-j2\pi i}{N} &= e^\frac{-j2\pi (i+N)}{N} \\
        g^{\frac{P-1}{N}(i+N)} &\equiv g^{\frac{P - 1}{N}} \mod{P}
\end{split}
\end{equation}
NTT can be written as the form in which $x[n]$ is a counterpoint to time domain in Fourier transform and $X[n]$ is a counterpoint to frequency domain in Fourier transform:
\begin{equation}
X[k] = \sum_{n=0}^{N-1}x[n]W_N^{kn}
\end{equation}
where $W_N^{i}=W_N^{i+N}=g^{\frac{P - 1}{N}i}$ is called the twiddle factors. 
All twiddle factors can be pre-computed and pre-loaded into memory instead of computing them on-the-fly to improve performance. This optimization can be apply to all seven algorithms.

\textbf{Decimation in Time (DIT)}:
NTT is a divide-and-conquer algorithm that can break down a N-point DFT transform into two smaller $\frac{N}{2}$-points transforms recursively, reducing the time complexity from $O(N^2)$ to $N\log{N}$. 
The possibility to divide-and-conquer is based on the following:
\begin{equation}
\begin{split}
    X[k] &= F_N(x[r])\\
    &= \Sigma^{\frac{N}{2}-1}_{r=0}x[2r]W^{rk}_{\frac{N}{2}} + W_N^k \Sigma^{\frac{N}{2}-1}_{r=0}x[2r+1]W^{rk}_{\frac{N}{2}}   \\
    &=F_{\frac{N}{2}}(x[2r]) + W_N^k F_{\frac{N}{2}}(x[2r+1])
\end{split}
\end{equation}
The binary representation of index can better describe the divide-and-conquer process,
$s$ indicates the current NTT stage. $A$, $B$ are composed of 0s and 1s, the length of $B$ is $s-1$, the length of $A$ is $width-s$, $C$ is either 1 or 0. 
$k=0bACB$ is the index of new representation in NTT:
\begin{equation}
\begin{split}
    X_s[k]
    &=X_x[0b\underbrace{\dots}_AC\underbrace{\dots}_B]\\
    &= X_s[0bACB]\\
    &=X_{s-1}[0bA0B] + W^{0bB}_{\frac{N}{2^{w-s}}} X_{s-1}[0bA1B]\\
    &=X_{s-1}[k] + W_{N}^{2^{w-s}\cdot 0bB} X_{s-1}[k + (1 << (s - 1))]
\end{split}
\end{equation}
NTTSuite implements the DIT using an iterative form instead of recursive form, as this is more suitable for HLS.

\begin{algorithm}
\caption{Decimation in Time}\label{alg:DIT}
    \begin{algorithmic}
        \REQUIRE $x[n] \textrm{ is input array}, W[n]\textrm{ is twiddle factor}, n = 2^L, \textrm{P is a prime}$
        \ENSURE $x \gets \textrm{DFT of x}$
        \STATE bit\_reverse(x)\\
        \FOR{$s \gets 1 \textrm{ to } L$}
            \STATE $m \gets (1 << i)$
            \FOR{$k \gets 0 \textrm{ to } n-1$}
                \STATE $tw \gets W[(1 << (L - s)) * k]$\\
                \FOR{$j \gets 0 \textrm{ to }  n \textrm{ by } m$}
                    \STATE $f_1 \gets x[j + k]$
                    \STATE $f_2 \gets (tw * x[j + k + (m >> 1)]) \% P$
                    \STATE $x[j + k] \gets (f_1 + f_2) \% P$
                    \STATE $x[j + k + (m >> 1)] \gets (f_1 - f_2) \% P$
                \ENDFOR
            \ENDFOR
        \ENDFOR
    \end{algorithmic}
\end{algorithm}

\textbf{Decimation in Frequency (DIF)}:
Unlike DIT, which does decimation in time domain, DIF decimates in the frequency domain.
The recursive form of DIF is:
\begin{equation}
    \begin{split}
    X[2k]
    &= \Sigma^{\frac{N}{2}-1}_{n=0}[x(n)+x(n+\frac{N}{2})]W^{nk}_{\frac{N}{2}}\\
    &=F_{\frac{N}{2}}[x(n)+x(n+\frac{N}{2})]\\
    X[2k+1]
    &= \Sigma^{\frac{N}{2}-1}_{n=0}[x(n)-x(n+\frac{N}{2})]W^n_N W^{nk}_{\frac{N}{2}}\\
    &=F_{\frac{N}{2}}[x(n)-x(n+\frac{N}{2})] W^n_N
\end{split}
\end{equation}
From the symmetrical expression of DIT and DIF, the major difference is butterfly operations between each stages and the order of bit-reverse.
\begin{algorithm}
\caption{Decimation in Frequency}\label{alg:DIF}
    \begin{algorithmic}
        \REQUIRE $x[n] \textrm{ is input array}, W[n]\textrm{ is twiddle factor}, n = 2^L, \textrm{P is a prime}$
        \ENSURE $x \gets \textrm{DFT of x}$
        \FOR{$s \gets L \textrm{ to } 1$}
            \STATE $m \gets (1 << i)$
            \FOR{$k \gets 0 \textrm{ to } n-1$}
                \STATE $tw \gets W[(1 << (L - s)) * k]$\\
                \FOR{$j \gets 0 \textrm{ to }  n \textrm{ by } m$}
                    \STATE $f_1 \gets x[j + k]$
                    \STATE $f_2 \gets x[j + k + (m >> 1)] \% P$
                    \STATE $x[j + k] \gets (f_1 + f_2) \% P$
                    \STATE $x[j + k + (m >> 1)] \gets (tw \cdot ((f_1 - f_2) \% P) \% P$
                \ENDFOR
            \ENDFOR
        \ENDFOR
        \STATE bit\_reverse(x)\\
    \end{algorithmic}
\end{algorithm}

\textbf{Pease}:
The work in Pease~\cite{pease} introduces a new factorization method, which can be represented by Algorithm \ref{alg:pease}:
\begin{equation}
    F_{2^t}=\{\Pi^t_{k-1}(I_{2^{c-1}}\otimes F_2 \otimes T_c)\}R^{2^t}_2
\end{equation}

\begin{algorithm}
\caption{Pease Alogrithm}\label{alg:pease}
    \begin{algorithmic}
        \REQUIRE $x[n] \textrm{ is input array}, W[n]\textrm{ is twiddle factor}, n = 2^L, \textrm{P is a prime}$
        \ENSURE $x \gets \textrm{DFT of x}$
        \STATE bit\_reverse(x)
        \FOR{$s \gets L \textrm{ to } 1$}
            \STATE $base \gets \sim (0xFFFFFFFF << (c - 1)))$
            \FOR{$r \gets 0 \textrm{ to } \frac{N}{2} - 1$}
                \STATE $f_1 \gets x[r << 1]$
                \STATE $f_2 \gets x[(r << 1) + 1]$
                \STATE $y[r] \gets (f_1 + f_2) \% P$
                \STATE $y[r + \frac{N}{2}] \gets (f_1 - f_2) \% P$
            \ENDFOR
            \STATE Swap(x,y)
        \ENDFOR
    \end{algorithmic}
\end{algorithm}

\textbf{Stockham}:
Both DIT and DIF algorithms need a bit-reverse operation, whose time complexity is $O(N\log{N})$ and 
results in additional memory accesses.
To access all memory twice, Stockham modifies the natural order NTT algorithm and uses two arrays to avoid bit-reverse operation. 

\textbf{Flat-NTT}: HLS may be unable to effectively loop unroll/pipeline if the iteration count is not fixed.
Therefore, in the DIT and DIF algorithms, there is a variation that flattens two inner loops to a single loop~\cite{flat}.

\textbf{Six-step}: The Six-step algorithm splits a large NTT into several smaller ones. Although time complexity remains the same, more computational overhead is introduced.
This has the same time complexity but introduces computational overhead. 
When the number of data points is large (e.g., 16384 data points), smaller blocks are made to fit into a cache, which is beneficial to CPU performance.

\textbf{Pease\_nc}: We also optimize the Pease algorithm to save time doing overhead copy work named Pease\_nc. The Pease\_nc swaps the input and output array after each computation loop instead of a copy. To enable more parallelism, it uses two types of butterfly operations in different stages. This way, the algorithm can both save time doing extra work copy work and support abilities of unrolling and scaling to utilize more hardware effectively.

\makeatletter
\newcommand{\rmnum}[1]{\romannumeral #1}
\newcommand{\Rmnum}[1]{\expandafter\@slowromancap\romannumeral #1@}
\makeatother

\section{Optimizations}
Catapult HLS tool offers two major optimization method: Pipelining and Loop unrolling. In addition, HLS can also inline code using pragmas to optimize. Ozcan and Aysu introduce and optimize with the above method~\cite{ozcan_aysu_2020}.  
In our paper, we focus on analyzing the memory access pattern to breakdown data dependencies, parallelizing the algorithm using the above optimization method, and utilizing the on-chip block ram (BRAM). We categorize them into three major optimizations to establish baseline designs competitive with the state-of-the-art. 
\rmnum{1}) \textit{Pipelining.} This strategy divides a loop iteration into multiple stages so that the next iteration of loop can start before previous iteration completes once all data needed for earlier stage is ready, therefore improving throughput and performance.  
Decreasing the time between executing loop iterations is an effective way to increase performance by fully utilizing the hardware resources and increasing throughput and performance. 
However, this is challenging in NTT due to data dependency and memory resource contention. 
After analyzing the memory access pattern, we decide to optimize accesses by reducing the iteration intervals (II).
In the Access Pattern Analysis, we show the effect of each algorithm's access pattern on pipelining. 
\rmnum{2}) \textit{Parallelism. } The basic structure of each NTTSuite benchmark can be described as a nested loop: the outer loop iterates over stages and the inner loop computes the $\frac{N}{2}$ butterfly operations. Since each butterfly operation is independent, we can parallel the computations using multiple butterfly cores simultaneously to decrease latency. 
The challenge is again the memory access pattern, as
some algorithms require memory index remapping after each stage. 
We show that the Pease algorithm is highly amenable to parallelism with our memory access pattern analysis.
\rmnum{3}) \textit{Operation optimization.} 
We currently use 32-bit primes.
We identify the modular reduction operation as a bottleneck of the NTT.
To optimize this, NTTSuite includes a recent optimization to perform reduction on FPGAs~\cite{heax}.
We apply our modular reduction methods, which can optimize all modular operations to reduce the time consumption and usage of hardware resources. 

\begin{figure}[t]
    \includegraphics[width=\linewidth]{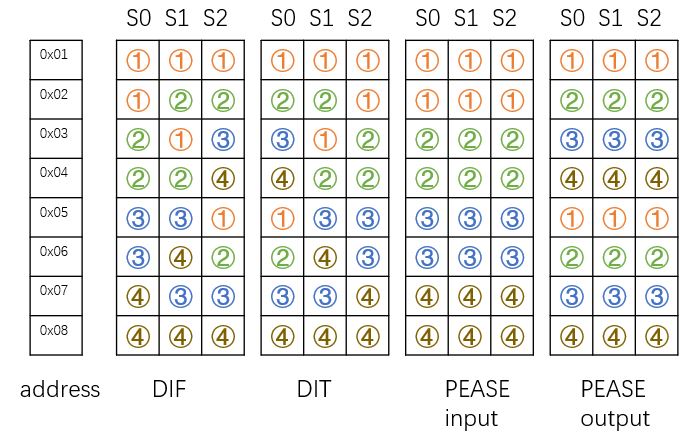}
    \vspace{-3em}
    \caption{Memory access pattern for the three major NTT types.}
    \label{fig:pattern}
    \vspace{-1em}
\end{figure}

\subsection{Access Pattern Analysis for Pipelining and Parallelism}

Though there are many ways to implement NTT, the major computation and memory access patterns can be categorized into three types: DIT, DIF, and Pease (i.e., constant geometry) shown in Algorithm ~\ref{alg:DIT}~\ref{alg:DIF}~\ref{alg:pease}.  
All other variations are designed to satisfy different situations like solving the extra bit reversal, loop flattening, or using swap to replace traditional copy work. 
Regardless, their computation paradigms are all in the three mentioned types.

\textbf{Pipelining.}
In NTTSuite, Algorithms DIT, DIF, and Peace are implemented in the same way: the outer loop iterates stages and the inner loop comprises $\frac{N}{2}$ non-overlapping butterfly operations. 
Operations in stage $n+1$ must wait until all operations in stage $n$ are complete and all inner loop butterflies have no data dependencies between each other.
Thus, it is possible to pipeline and unroll onto the inner loop. 
In the access pattern graph, see Figure~\ref{fig:pattern}, \textcircled{i} denotes the two inputs needed to compute the $i$-th butterfly operation in the inner loop and S$j$ indicates the stage $j$.
To fully pipeline DIT, \textcircled{1} and \textcircled{2} should be in different \textit{blocks}, i.e., memory partitions.
The challenge with the DIT algorithm is that the butterfly input pattern changes for every stage, inevitably resulting in block/partition conflicts, i.e., structural hazards, that limit the opportunity for pipelining.
E.g., if we partition memory for perfect pipelining in S0, Figure~\ref{fig:pattern} shows that stages S1 and S2 will incur conflicts due to the pattern the data was written back.
This means address 0x00 must be in different blocks with 0x01, 0x02, 0x03, and 0x06, which is not realizable in our memory layout.
Since the DIF algorithm's memory access is like DIT algorithm's memory access, which also has the stride pattern, we found the HLS tool could not produce well-pipelined designs with these algorithms.

As seen in Figure~\ref{fig:pattern}, the the Pease algorithm input follows a simple memory access pattern that we can easily separate \textcircled{1}\textcircled{2}, \textcircled{2}\textcircled{3},\textcircled{3}\textcircled{4} by setting 0x01, 0x02, 0x05, 0x06 to a block and other address to the other block in all stages. 
(The same works for the output array.) 
Furthermore, because our Pease implementation is out of place and the butterfly operations go from one input array to another output array, the resource contention is also much lower than DIT and DIF.
Leveraging the FPGA's of Dual port memories, the Pease algorithm can have $\frac{N}{2}$ read ports resource and $\frac{N}{2}$ write ports resource in $\frac{N}{2}$. This enables us to achieve a pipeline initiation interval of 1 (II=1), even without memory partitioning.

\textbf{Parallelism.}
We further improve performance using parallel hardware to execute multiple butterflies from the same stage simultaneously. 
In DIF algorithm, if making parallel \textcircled{1}\textcircled{2}\textcircled{3}\textcircled{4} in stage 1, then \{0x01, 0x02\} and \{0x03, 0x04\} and \{0x05, 0x06\} and \{0x07, 0x08\} must be separated. 
But in stage 2 we have: \{0x01, 0x03\}, \{0x02, 0x04\}, \{0x05, 0x07\}, \{0x06, 0x08\}, which means all eight addresses must be separated. 
(The same holds for DIT algorithm.)
For the Pease algorithm, we can simply divide the input into four parts by setting blocksize=2 and divide the output into four parts by setting interleave=4 in HLS.
(Blocksize=B divides memory into multiple B-word memory partitions. Interleave=M places adjacent memory locations into different memory partitions. E.g., interleave=4 would partition memory 0,4 into the first memory block, 1,5 into the second, 2,6 into the third, and 3,7 into the fourth.)
Computing four parallel butterflies on the 8-point problem, there is no difference among these algorithms because DIT, DIF, and Pease all use 8 BRAMS.
However, with the same analysis on arbitrary $2^t$ points, DIF and DIT must have $2^t$ BRAMS to do 4-butterflies in parallel, while Pease still needs only 4 blocks on its input and output array. 
Thus, we find Pease is amenable to parallelism and can scale up or down depending on constraints.
It is also possible to do memory remapping after each stage to support parallelism and pipelining for DIT and DIF algorithms, but it also introduces resource overhead and design complexity. 
In general, the Pease algorithm has the best quality to do pipeline and parallel execution.

\subsection{No Copy Optimization}
In the following stage, the current input array would now be the output array, which needs to be written in a different memory access pattern not the same as before to be read. 
We find that the memory swap costs can be eliminated for the Pease algorithm.
As above, when using dual port memory, both the input and output array can provide sufficient resources to realize pipelining. So the ability to pipeline effectively is not changed after a swap. To utilize parallelism, we can set interleave=4 for both input and output, unlike what we did in Pease. 
Then for arbitrary $2^t$ data points, in each iteration, input accesses to memory addresses \{$8r+0$, $8r+1$, $8r+2$, $8r+3$,$8r+4$, $8r+5$, $8r+6$, $8r+7$,\} while output accesses to memory addresses \{$4r+0$, $4r+1$, $4r+2$, $4r+3$,$4r+2^{t-1}$, $4r+2^{t-1}$, $4r+2^{t-1}$, $4r+2^{t-1}$\}. 
Also, from the dual port memory all memory addresses can be accessed simultaneously. That means we can simply set Interleave=N, where N is the number of butterfly units used in parallel. Thus, the input and output arrays can be set up with the exact same memory partition making it safe and effective to swap them while still maintaining pipeline and parallelism optimizations.

To reproduce our optimized Pease\_nc (16 butterfly units) result, follow the steps below.
First, configure the memory resource type to dual-port RAM, which best supports pipelining and parallelism. 
After that, set memory partitions of both input and output arrays to interleave=16, allowing at most 16 butterfly cores to work together. 
Finally, set pipeline interleave to one (II=1), which can significantly increase throughput. 

In the end, we can run our Pease\_nc on 4096 inputs with a latency of 8.6us shown in the result.
HEAX, a state-of-art FPGA implementation for HE, reports a 4096 point NTT latency of 11us, using more hardware resources~\cite{heax}.
Thus, the Pease\_nc optimization provides a 30\% speedup.

\subsection{Modular Reduction}
All data operations in NTT are modular with respect to a prime $P$.
Initially, we specified the operations and implicitly let HLS synthesize the hardware design.
We noticed that hardware allocated for modular reduction was inefficient.
Therefore, all NTT algorithms in NTTSuite use an explicit implementation for modular reduction following the algorithm proposed by HEAX~\cite{heax}.
As shown in the results section, this optimization has a significant performance impact of 12$\times$.

All $mod$ operations in computation can be implemented by several judging, shift, plus and minus operations instead of taking remainder of modulo. 
The mod are most common operations in NTT algorithms, to fully optimize it can significantly relax the critical path in FPGA design which can also increase frequency limitation satisfying the slack requirements. 
From the fact that $a<P \textrm{ and } b<P$, we have $0<(a+b)<2P$ and $-P<(a-b)<P$ which means all $(a+b)\mod{P} \ textrm{and} (a-b)\mod{P}$ can be realized by an assert and an plus or subtraction.
All the multiplication happens between a twiddle factor and a variable $x$, so it is possible to pre-compute all $tw_h[i]=\lfloor (y<<w) / m \rfloor$ as an array for twiddle factors, which can be used in Modular Reduction\cite{modularReduction}.

\begin{algorithm}
\caption{modulo\_add}\label{alg:moduloAdd}
    \begin{algorithmic}
        \REQUIRE $base, m$
        \STATE $result \gets base \% m$
        \IF{$result > m$}
            \RETURN $result - m$
        \ELSIF{$result < 0$}
            \RETURN $result + m$
        \ELSE
            \RETURN $result$
        \ENDIF
    \end{algorithmic}
\end{algorithm}

\begin{algorithm}
\caption{modulo\_mult}\label{alg:moduloMult}
    \begin{algorithmic}
        \REQUIRE $x, m, tw, tw_h=\lfloor (y<<w) / m \rfloor$
        \ENSURE $z = x \cdot tw \mod{P}$
        \STATE $z_a \gets (x \cdot y) \quad \& \quad \textrm{0xFFFFFFFF}$
        \STATE $z_b \gets (((x \cdot tw_h) >> w)\cdot p) \quad \& \quad \textrm{0xFFFFFFFF}$
        \STATE $z \gets z_a - z_b$
        \IF{$z \leq 0$}
            \RETURN $z - m$
        \ELSE
            \RETURN $z$
        \ENDIF
    \end{algorithmic}
\end{algorithm}
\section{Performance and Analysis}
In this section, we show how our optimization methods improve resource usage and performance. 
In addition, we show how different algorithms perform on CPU (Intel E5), GPU (RTX 8000), and FPGA (Xilinx v7690t1761-2).  
For a thorough evaluation, we evaluate all NTTSuite algorithms on three input sizes: 1024, 4096, and 16834.

\subsection{Methodology}
To fairly compare the acceleration performance between GPU and FPGA, we calculate the time after we copy the input vectors to the device and before we copy the vectors from the device back to the host memory.
For the GPU, to get a more precise and stable result, we profile the benchmarks 100 times each and report the mean as the measured GPU computation time.

For FPGA experiments, we use Catapult HLS  version 10.5c to generate RTL and reports including the throughput (cycles and time), latency (cycles and time), total area, and slack, which we used as our final time result. 
Waveforms are generated using QuestaSim for RTL simulation and verification.
The generated RTL is then imported to Vivado Design Suite version 2019.1 and synthesized on the FPGA chip. 
Vivado Design Suite reports the hardware resources used by each algorithm, including LUTs, FFs, BRAMs, and DSPs. Since the input and output vector arrays map to memory ports, those BRAMs are not included. We report each algorithm's best-optimized results to make comparisons using the best-performing designs.

\begin{table}[t]
\caption{NTTSuite FPGA results using various vector sizes.
        L, F, D, and B stand for LUTs, FFs, DSPs, and BRAMs, respectively.}
\vspace{-1em}
\begin{tabular}{|c|c|r|c|c|c|c|c|}
\hline
Name      & Size & Time(us) & Freq & L   & F    & D & B \\ \hline
DIF        & 1K   & 851.57   & 100  & 973   & 536   & 11  & 0    \\ \cline{2-8} 
           & 4K   & 4009.41  & 100  & 1241  & 694   & 12  & 0    \\ \cline{2-8} 
           & 16K  & 18601.63 & 100  & 1388  & 680   & 20  & 0    \\ \hline
DIT        & 1K   & 1057.97  & 100  & 2248  & 998   & 5   & 0    \\ \cline{2-8} 
           & 4K   & 5048.13  & 100  & 1076  & 671   & 11  & 0    \\ \cline{2-8} 
           & 16K  & 23198.11 & 100  & 1916  & 776   & 20  & 0    \\ \hline
Flat-NTT   & 1K   & 98.73    & 167  & 1142  & 446   & 11  & 0    \\ \cline{2-8} 
           & 4K   & 466.88   & 167  & 1254  & 470   & 11  & 0    \\ \cline{2-8} 
           & 16K  & 2159.92  & 167  & 1306  & 494   & 11  & 0    \\ \hline
Pease      & 1K   & 5.27     & 200  & 20445 & 22313 & 32  & 320  \\ \cline{2-8} 
           & 4K   & 21.83    & 200  & 20350 & 22349 & 32  & 320  \\ \cline{2-8} 
           & 16K  & 97.19    & 200  & 19203 & 22364 & 32  & 320  \\ \hline
Pease\_nc  & 1K   & 7.18     & 196  & 6005  & 8592  & 4   & 40   \\ \cline{2-8} 
(4cores)   & 4K   & 31.93    & 196  & 6079  & 6198  & 4   & 40   \\ \cline{2-8} 
           & 16K  & 146.96   & 196  & 6073  & 5334  & 4   & 40   \\ \cline{2-8} 
           & 64K  & 669.30   & 196  & 6145  & 5409  & 64  & 40   \\ \hline
Pease\_nc  & 1K   & 2.27     & 196  & 23474 & 32866 & 16  & 160  \\ \cline{2-8} 
(16cores) & 4K   & 8.60     & 196  & 23696 & 32976 & 16  & 160  \\ \cline{2-8} 
           & 16K  & 37.50    & 196  & 23737 & 27902 & 16  & 160  \\ \hline
Six-step  & 1K   & 13.00    & 150  & 28816 & 8930  & 82  & 32   \\ \cline{2-8} 
           & 4K   & 38.76    & 150  & 47038 & 19453 & 163 & 64   \\ \cline{2-8} 
           & 16K  & 144.80   & 150  & 77558 & 26359 & 163 & 128  \\ \hline
Stockham   & 1K   & 114.31   & 135  & 1602  & 1329  & 10  & 8    \\ \cline{2-8} 
           & 4K   & 546.92   & 135  & 1659  & 1351  & 10  & 8    \\ \cline{2-8} 
           & 16K  & 2550.22  & 135  & 1741  & 1373  & 10  & 16   \\ \hline
Heax~\cite{heax}
           & 4K   & 11   & 275  & N/A  & N/A  & 1185  & 1731    \\ \hline
\end{tabular}
\label{table1}
\vspace{-2em}
\end{table}

\subsection{Optimization Analysis}

Since the optimized Pease algorithm yields the fastest speedup and is most suitable for all types of optimization methods, we take this algorithm as an example to introduce and illustrate the effects of the various optimization methods supported in NTTSuite.
After the techniques are applied to the pease no-copy algorithm, the performance is improved based on different optimization methods, as Figure \ref{fig:optm_applied} suggests.
Note that in Figure~\ref{fig:optm_applied}
optimizations accumulate from left to right to show how much speedup each provides.

In addition, Figure~\ref{fig:optimizations} reports the resources used by each optimization approach on the Peace No-copy algorithm. 
Results show the modular reduction optimization significantly reduces the number of LUTs and FFs used compared to the original algorithm, which suggests this optimization method can both release the pressure of resource usage and decrease the latency time for acceleration.
Pipelining, based on Figure~\ref{fig:optimizations}, increases the usage of FFs significantly and the usage of DSPs slightly in order to improve performance, compared to the original and Modular algorithm in Figure~\ref{fig:optm_applied}. 
This is because the pipeline requires slightly more hardware to implement but can drastically improve performance by better utilizing allocated resources, which we observe as the resources only increase modestly while speedup increases from 12.17 to 78.19.

We also explored memory partitioning by increasing interleaving (or decreasing block size) combined with unrolling loop to extract more parallelism.
Memory interleave separate adjacent data stored in same BRAM to two or more BRAMs that are not adjacent, allowing memory access to different block RAMs and computation modules to perform simultaneously in same clock cycles. This optimization requires more resources (for the parallel hardware) but significantly improves performance, see Figure~\ref{fig:optm_applied}.

\begin{figure}[t]
        \vspace{2em}
	\centering
	\includegraphics[width=\columnwidth]{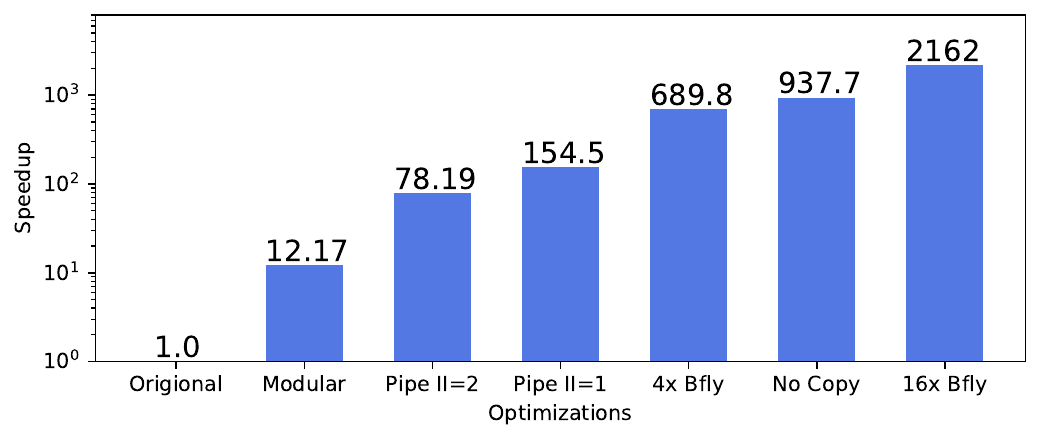}
	\caption{Optimizations applied cumulatively to the Pease\_nc algorithm. Speedups are noted on top of the bars.}
        \label{fig:optm_applied}
        \vspace{-1em}
\end{figure}

\begin{figure}[t]
	\centering
	\includegraphics[width=\columnwidth]{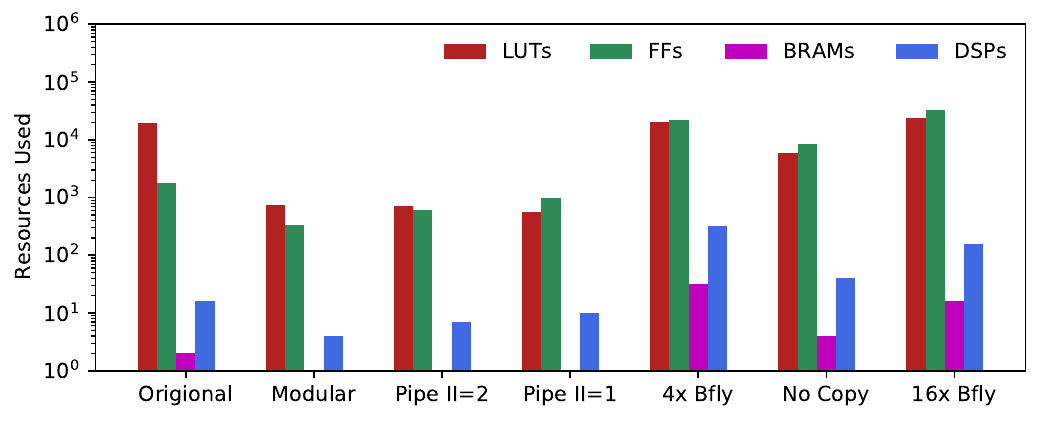}
	\vspace{-2.5em}
	\caption{Optimization resource utilization for the Pease\_nc algorithm.}
	\label{fig:optimizations}
\end{figure}

\begin{figure}[t]
	\centering
	\includegraphics[width=\columnwidth]{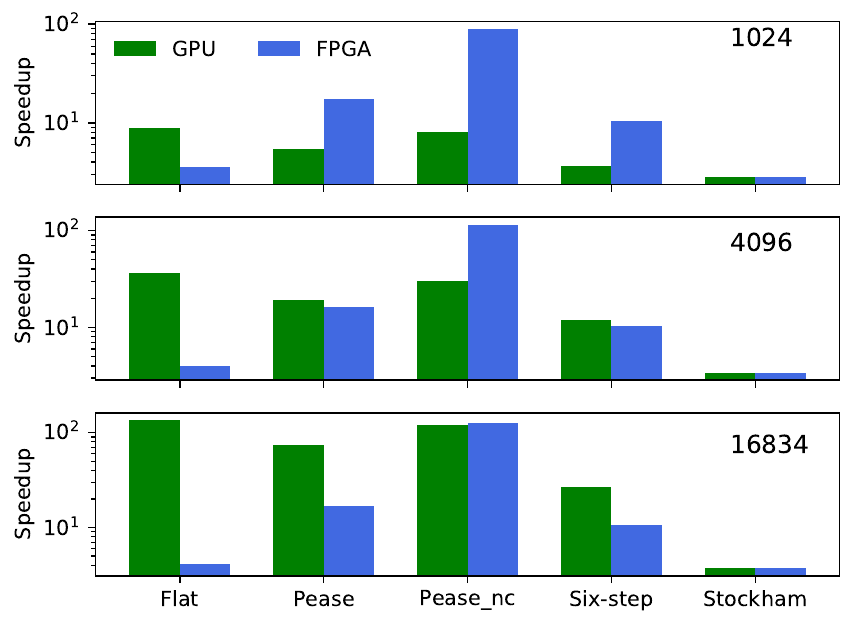}
	\caption{GPU and FPGA speedup relative to the CPU. Problem size is noted in the top right of each plot.}
	\label{fig:all_speedup}
\end{figure}

\begin{table}[t]
\caption{NTTSuite Pease No-copy FPGA Post-implementation Resource Utilization}
\centering
\vspace{-1em}
\begin{tabular}{|c|c|c|c|}
\hline
Resource   & Utilization & Available & Utilization \\ \hline
LUT & 9829 & 663360 & 1.48    \\ \hline
LUTRAM & 547 & 293760 & 0.19 \\ \hline
FF & 10511 & 1326720 & 0.79 \\ \hline
BRAM & 105 & 2160 & 4.86 \\ \hline
DSP & 11 & 5520 & 0.2 \\ \hline
IO & 4 & 702 & 0.57 \\ \hline
GT & 1 & 64 & 1.56 \\ \hline
BUFG & 5 & 1248 & 0.40 \\ \hline
PCIe & 1 & 6 & 16.67 \\ \hline
\end{tabular}
\label{table1}
\vspace{2em}
\end{table}

\begin{figure}[t]
   \centering
   \includegraphics[width=\columnwidth]{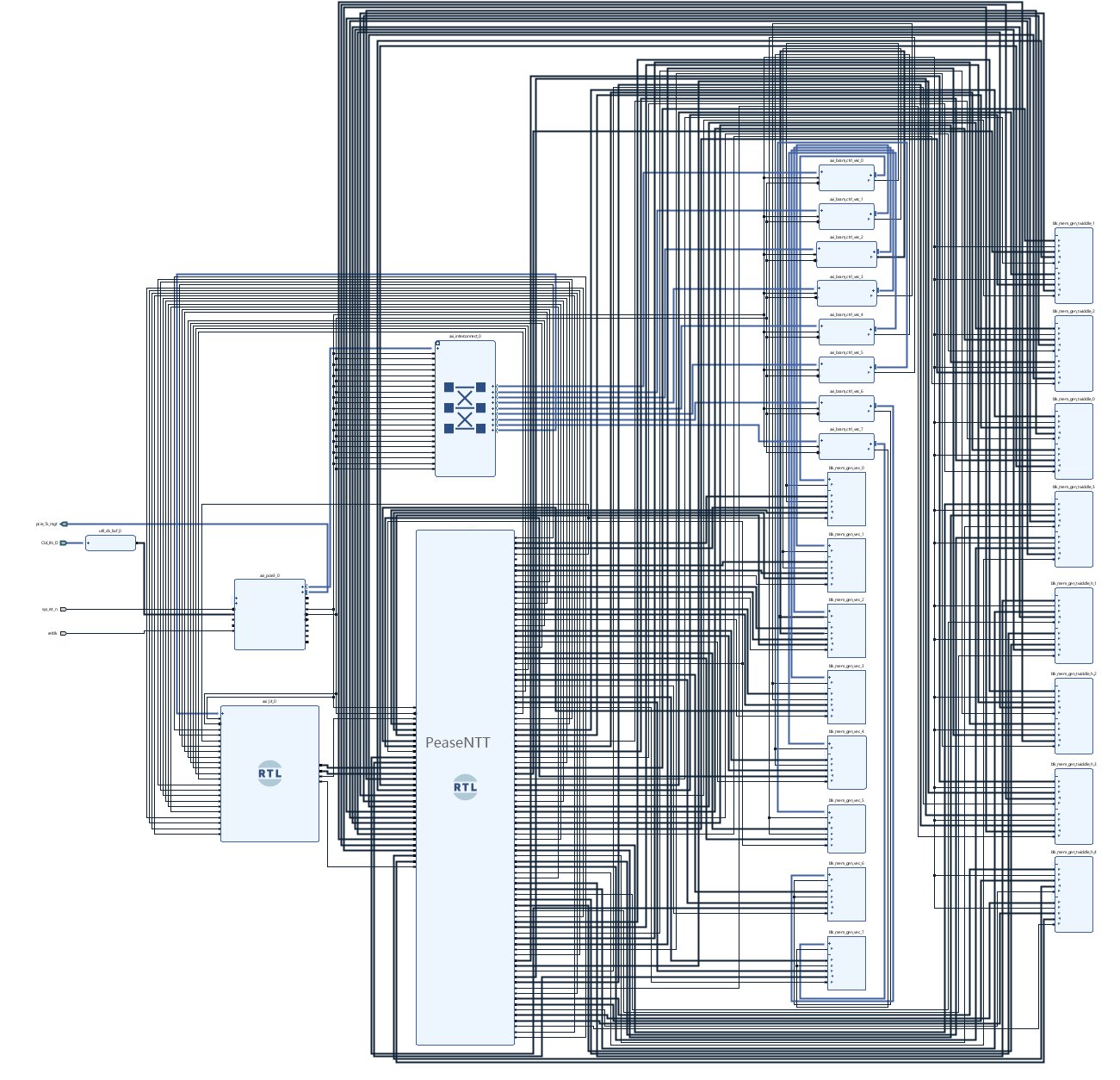}
   \caption{Block Diagram of Pease No-copy with AXI interconnect, BRAM modules, and PCIe-AXI Bridge}
   \label{fig:all_speedup}
\end{figure}

\subsection{Observation and Analysis}
Figure \ref{fig:all_speedup} compares NTTSuite algorithms running on three vector size on Flat, Pease, Pease No-copy, Six-step, and Stockham algorithms (which are improvement of traditional DIT and DIF). 
Each bar shows FPGA (blue) or GPU (green) speedup normalized to the CPU runtime.
We find that that the Stockham Algorithm performs better on GPU compared to FPGA. Pease, Pease\_nc, and Six-step perform better on FPGA. 
Although DIF and DIT perform better on GPU, results on FPGA are much slower than running on CPU. 
According to the result in Table~\ref{table1}, in each algorithm, as the vector size increases, the latency time increases proportionally to the vector size's increment.
The Pease\_nc algorithm is by far the fastest algorithm on the FPGA.
To achieve this performance, it also consumes the most hardware resource compared to the hardware resource consumed by other algorithms on  FPGA. 

Both DIT and DIF receive few or even negative speedups on FPGA and GPU due to the memory access pattern in which more data dependency exists. The nested loop structure of DIT and DIF leads to difficulties for HLS to unroll and pipeline loops since iterations are not fixed and complex dependencies cannot be resolved. Since the FPGA usually has a lower clock frequency, poor performance is expected. A similar issue arises with our GPU implementation since the nested loop will lead to extra function calls on GPU. Flatting them easily resolves the issues, just like the Flat-NTT algorithm did. In this case, a powerful multi-core CPU is more cost-efficient since instruction-level parallelism resolves the data dependency issue better than utilizing SIMD in GPU and the scratch-pad block memory in FPGA.

Flat, Pease, Pease No-copy, Six-step, and Stockham algorithms receive huge boost from FPGA while DIT and DIF runs slower than CPU. The FPGA platform has unique hardware resource including the LUT, LUTRAM, BRAM, FF, and DSP while GPU and CPU do not. LUTs and DSPs facilitate parallel computation while large BRAM and LUTRAM can be used as scratch pad memory to reduce data access time and number of data movement comparing to multi-level caches in CPUs and GPUs. In addition, during the synthesis process, computation logic is further optimized for speed which also contributes to the overall speedup.

\section{Related Work}

Previous works~\cite{CPU_Microsoft2016, CPU_Seiler18, HEXL2021} focus on optimizing the NTT algorithm itself as well as modular reduction~\cite{Barret_1987, MR_Yanik2002}.
Other focus on Lattice-based computations directly~\cite{Longa2016}.

In our work, we focus on six algorithms: Decimation-in-Time (DIT~\cite{Cooley1965}), Decimation-in-Frequency (DIF~\cite{DIF}), Flat-NTT~\cite{flat}, Pease Algorithm~\cite{pease}, Stockham~\cite{Cochran1967}), and Six-step~\cite{sixstep} that could be applied in any research using NTT. Other works~\cite{GPU_Ozgun2021, SHEonGPU_Badawi2018} propose GPU accelerated solutions. Kim, Jung, Park, and Ahn~\cite{Kim2020} compare different algorithms on GPU and analyze their performance and limitations at the same time. Several works~\cite{FPGA_HE_Sinha2019, Nejatollahi2020, FPGA_Ye2021} present FPGA solution that focus on optimizing algorithms and implementation using verilog. Recent work~\cite{Nejatollahi2020} also present optimized solution using Vivado HLS and synthesis on FPGA. 
Many others are actively working to accelerate the entire HE computation~\cite{samardzic2021f1, cheetah, soni2023rpu, samardzic2022craterlake, kim2022bts}.


In summary, while there have been many prior works on accelerating NTT, to the best of our knowledge there are no common set of implementations that evaluates on all three platform: CPU, GPU, and FPGA. NTTSuite aims to fill this gap by providing open-source implementations and optimizations.

\section{Conclusion}
This paper develops NTTSuite: a collection of seven NTT algorithms with CPU, GPU, and HLS implementations.
The benchmarks include testing infrastructure and performance optimizations to enable researchers to
devise novel hardware primitives for NTT.
NTT algorithm performance can be seen as a core bottleneck in homomorphic encryption, one of the foremost technologies to enable true privacy-preserving computation. 
In future work, we intend to extend NTTSuite to include more HE primitives
and research new hardware structures and optimizations to realize HE using accelerators.

Now all our works are simulated on software but not deployed on an actual FPGA chip. So in the next stage, we will apply our NTT algorithms to configure the netlist after synthesizing onto the hardware and do timing simulation to testify the design.


\section{Acknowledgements}
This work was supported in part by the Applications Driving Architectures (ADA) Research Center, a JUMP Center co-sponsored by SRC and DARPA.
The views, opinions and/or findings expressed are those of the author and should not be interpreted as representing the official views or policies of the Department of Defense or the U.S. Government.

\bibliographystyle{IEEEtran}
\bibliography{ref}

\end{document}